\documentclass[aps,prl,twocolumn,groupedaddress,showpacs]{revtex4} 

\usepackage{graphicx}
\usepackage{dcolumn}
\usepackage{bm}

\begin{document} 

\title{Thermal Conductivity of Pr$_{1.3-x}$La$_{0.7}$Ce$_x$CuO$_4$ 
Single Crystals and Signatures of Stripes in an Electron-Doped Cuprate} 

\author{X. F. Sun} 
\author{Y. Kurita} 
\altaffiliation{also at Department of Physics, Tokyo University of Science, 
Shinjuku-ku, Tokyo 162-8601, Japan.} 
\author{T. Suzuki} 
\altaffiliation{also at Department of Physics, Tokyo University of Science, 
Shinjuku-ku, Tokyo 162-8601, Japan.} 
\author{Seiki Komiya} 
\author{Yoichi Ando} 
\email{ando@criepi.denken.or.jp} 
\affiliation{Central Research Institute of Electric Power Industry, 
Komae, Tokyo 201-8511, Japan.} 

\date{\today} 

\begin{abstract} 

It was recently demonstrated that the anisotropic phonon heat 
transport behavior is a good probe of the stripe formation in 
La$_{2-x}$Sr$_x$CuO$_4$ (LSCO) [X. F. Sun {\it et al.}, Phys. 
Rev. B {\bf 67}, 104503 (2003)]. Using this probe, we examined an 
electron-doped cuprate Pr$_{1.3-x}$La$_{0.7}$Ce$_x$CuO$_4$ 
(PLCCO) and found that essentially the same features as those in 
LSCO are observed. Moreover, the in-plane resistivity $\rho_{ab}$ 
of lightly-doped PLCCO shows metallic behavior ($d\rho_{ab}/dT > 
0$) in the N\'eel ordered state with a mobility comparable to 
that in LSCO. It is discussed that these peculiar properties in 
common with LSCO signify the existence of stripes in 
electron-doped cuprates.

\end{abstract} 


\maketitle 

High-$T_c$ superconductivity shows up when either holes or 
electrons are doped to parent Mott-insulating cuprates. The 
essential features of the phase diagram, such as the N\'eel 
ordering in the lightly-doped regime and superconductivity in the 
moderately-doped regime, are approximately symmetric 
\cite{Takagi,Kastner} for hole- and electron-doped sides, with the 
undoped Mott insulator sitting in the middle, and this leads to a 
key paradigm in the study of the high-$T_c$ cuprates: the 
electron-hole symmetry. However, there are also disparities 
between the two sides, and understanding the similarities and 
differences between the hole- and electron-doped cuprates would be 
crucial for elucidating the fundamental nature of these materials. 
Of particular interest is whether the ``stripes" \cite{Carlson} (a 
sort of quasi-one-dimensional charge/spin density wave) exist in 
the electron-doped cuprates \cite{Sadori}, since such 
self-organized structures have been observed in the hole-doped 
cuprates \cite{stripes} and are discussed to be relevant to the 
occurrence of superconductivity \cite{Carlson}.

Most convincing evidence for the stripes in the hole-doped 
cuprates had been obtained by neutron scattering in the form of 
``incommensurate" superstructure peaks 
\cite{Tranquada,YamadaLSCO,Matsuda,Mook}; on the other hand, 
neutron scattering experiments on an electron-doped material 
Nd$_{2-x}$Ce$_x$CuO$_4$ (NCCO) found only commensurate magnetic 
peaks to coexist with the superconductivity \cite{YamadaNCCO}. 
Such commensurate peaks would naturally suggest that there is no 
superstructure in the spin system; however, if the stripes are 
``in-phase" domain boundaries of the spin system that leave 
strong commensurate peaks (as opposed to antiphase boundaries 
that give rise to only incommensurate peaks), the neutron data on 
NCCO can be consistent with such stripe structures 
\cite{Kivelson}. Thus, it is still an open question whether the 
stripes exist in the electron-doped cuprates. Here we show that 
the transport properties of an electron-doped cuprate demonstrate 
peculiar features that are similar to those in the hole-doped 
cuprates, where such features \cite{mobility,Sun1} have been 
shown to be naturally understood as consequences of the stripe 
formation. Therefore, it is most likely that the stripes exist in 
electron-doped cuprates and that there is an approximate 
electron-hole symmetry not only for the occurrence of 
superconductivity but also for the stripe formation.

Recently, we have demonstrated \cite{Sun1} that the $c$-axis 
phonon heat transport is a good probe of the stripe formation in 
lightly hole-doped La$_{2-x}$Sr$_x$CuO$_4$ (LSCO); namely, the 
spin stripes in this system are well-ordered in the CuO$_2$ 
planes but are disordered along the $c$ axis \cite{Matsuda}, 
which (perhaps through the strong spin-lattice coupling 
\cite{Lavrov}) causes the $c$-axis phonons to be strongly 
scattered. We have also demonstrated \cite{mobility} that the 
in-plane resistivity $\rho_{ab}$ of lightly-doped LSCO crystals 
shows metallic behavior ($d\rho_{ab}/dT > 0$) even in the 
long-range-ordered N\'eel state, where the hole mobility is 
virtually the same as that in optimally-doped samples; such an 
unusual metallic behavior can naturally be understood if doped 
holes form self-organized ``rivers" whose distance changes with 
doping \cite{mobility,Trieste}. Taking these features as 
signatures of stripes, we set out to examine the transport 
properties of (Pr,La)$_{2-x}$Ce$_x$CuO$_4$ (PLCCO) system, for 
which Fujita {\it et al.} reported \cite{Fujita} that 
high-quality single crystals can be grown by the 
traveling-solvent floating-zone (TSFZ) technique. We choose the 
composition of Pr$_{1.3-x}$La$_{0.7}$Ce$_x$CuO$_4$, which is 
structurally stable in the lightly Ce-doped region \cite{Ikeda}.

High-quality Pr$_{1.3-x}$La$_{0.7}$Ce$_x$CuO$_4$ single crystals 
($x$ = 0, 0.01, 0.03, 0.05, 0.08, 0.10, and 0.13) are grown by the 
TSFZ technique in flowing oxygen. The crystals are cut into 
rectangular platelets with a typical size of $2.5 \times 0.5 
\times 0.1$ mm$^3$, where the $c$ axis is perpendicular or 
parallel to the platelets within an accuracy of better than 
1$^{\circ}$. The $ab$-plane and $c$-axis thermal conductivities 
($\kappa_{ab}$ and $\kappa_c$) are measured by a steady-state 
technique \cite{Sun1}. Several samples are measured for each Ce 
concentration to check for the reproducibility, resulting in the 
uncertainties in the reported values of less than 10\%. Standard ac 
four-probe method is employed to measure the resistivity 
\cite{LSCO_MR}. Note that the properties of the electron-doped 
cuprates are very sensitive to oxygen non-stoichiometry 
\cite{Schultz,Jiang}; reduction annealing is necessary to remove 
excess (apical) oxygen from oxygenated (as-grown) crystals and 
to introduce electrons into the CuO$_2$ planes. For this work, to 
remove sufficient apical oxygens while introducing only a small 
amount of oxygen vacancies in the reduced samples, we performed 
thermogravimetry analyses and settled on the annealing at 
850--875$^{\circ}$C for 24 h under Ar gas flow \cite{note}. With 
this condition, the difference in the oxygen content between the 
oxygenated and reduced samples (measured by the weight change 
after the Ar-annealing) is rather small (for example, the 
difference is 0.1\% for the $x$ = 0.05 sample).

\begin{figure}  
\includegraphics[clip,width=8.5cm]{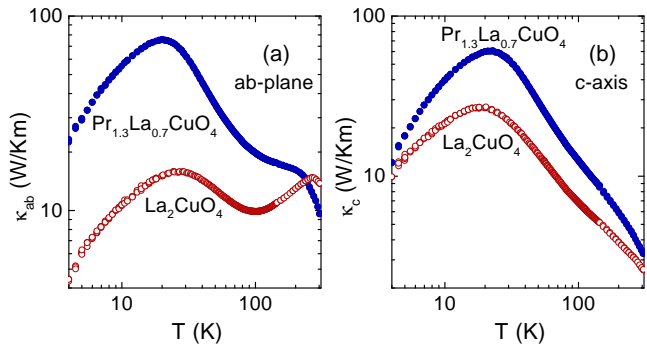} 
\caption{Thermal conductivity of parent cuprates 
Pr$_{1.3}$La$_{0.7}$CuO$_4$ and La$_2$CuO$_4$ (taken from Ref. 
\cite{Sun1}) along (a) the $ab$ plane and (b) the $c$ axis.} 
\end{figure} 

To begin with, let us discuss the temperature dependences of 
$\kappa_{ab}$ and $\kappa_c$ of undoped 
Pr$_{1.3}$La$_{0.7}$CuO$_4$ (PLCO) crystals shown in Fig. 1, where 
the data for La$_2$CuO$_4$ (LCO) \cite{Sun1} are also included for 
comparison. The LCO crystals exhibit a pronounced phonon peak at 
low temperature in both $\kappa_{ab}$ and $\kappa_c$, while at 
high temperature close to the N\'eel temperature $T_N$ ($\simeq$ 
300 K) another peak appears in $\kappa_{ab}$ and this has been 
attributed to the magnon heat transport 
\cite{Sun1,Nakamura,Hess1}. One can see that PLCO shows much 
higher low-$T$ peak in both $\kappa_{ab}$ and $\kappa_c$, 
suggesting an intrinsically better phonon transport in PLCO 
\cite{Cohn}; also, these data testify the high quality of our 
crystals. In the $\kappa_{ab}(T)$ data, one can further see that 
the magnon contribution to the heat transport is evident in PLCO, 
where the larger phonon contribution turns what was a clear peak 
in LCO into a hump; note that the position of this hump is 
consistent with $T_N (\simeq$ 260 K in Pr$_2$CuO$_4$ \cite{Cox}). 
It is fair to note that this is not the first time a signature of 
the magnon heat transport in the so-called $T'$-phase is observed, since 
similar hump profile was reported for polycrystalline 
Pr$_2$CuO$_4$ samples \cite{Sologubenko}. However, only by using 
single crystals can one confirm that the hump appears only in the 
$ab$ plane, which is essential for attributing this hump to the 
magnons in two-dimensional (2D) spin systems.

\begin{figure} 
\includegraphics[clip,width=6.5cm]{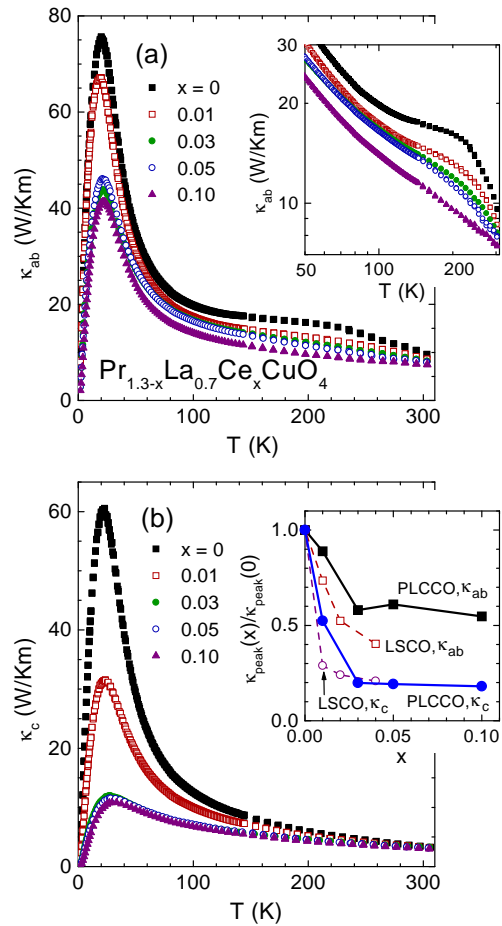} 
\caption{Thermal conductivity of reduced single crystals of 
Pr$_{1.3-x}$La$_{0.7}$Ce$_x$CuO$_4$ along (a) the $ab$ plane and 
(b) the $c$ axis. Insets: (a) $\kappa_{ab}(T)$ data near the 
high-$T$ hump in a log-log plot; (b) Doping dependences of the 
peak values of $\kappa_{ab}$ and $\kappa_c$, normalized by the 
values at $x$ = 0, for PLCCO and LSCO (data for LSCO are taken 
from Ref. \cite{Sun1}).} 
\end{figure} 

Figure 2 shows how $\kappa_{ab}(T)$ and $\kappa_c(T)$ of PLCCO 
change with doping. With increasing $x$, the high-$T$ hump in 
$\kappa_{ab}$ becomes weaker and almost disappears at $x$ = 0.05 
[which can be better seen in the inset to Fig. 2(a)], even though 
the N\'eel temperature shows just a weak $x$-dependence up to $x$ 
= 0.08 in PLCCO \cite{Fujita}. This behavior is somewhat similar 
to that in LSCO, where the magnon peak is completely suppressed 
with only 1\%-Sr-doping, for which $T_N$ is still as high as 240 K 
\cite{Sun1}.

The main finding in Fig. 2 is that the low-$T$ phonon peak shows 
very anisotropic evolution upon electron doping; namely, the 
low-$T$ peak of $\kappa_c$ is dramatically suppressed with slight 
Ce doping, while the doping effect on $\kappa_{ab}$ is much more 
modest. The inset to Fig. 2(b) shows a plot of the $x$-dependence 
of the peak height relative to that of $x$ = 0; here, similar data 
for LSCO \cite{Sun1} are also shown. One can see that the drastic 
suppression of the $c$-axis phonon peak upon slight carrier doping 
is common to LSCO and PLCCO. Note that the contribution of 
electronic heat transport is negligibly small at low temperatures 
in these insulating crystals \cite{Sun1}, so the doping dependence 
in the low-$T$ heat transport is solely due to some doping-induced 
scattering of phonons.

\begin{figure} 
\includegraphics[clip,width=8.5cm]{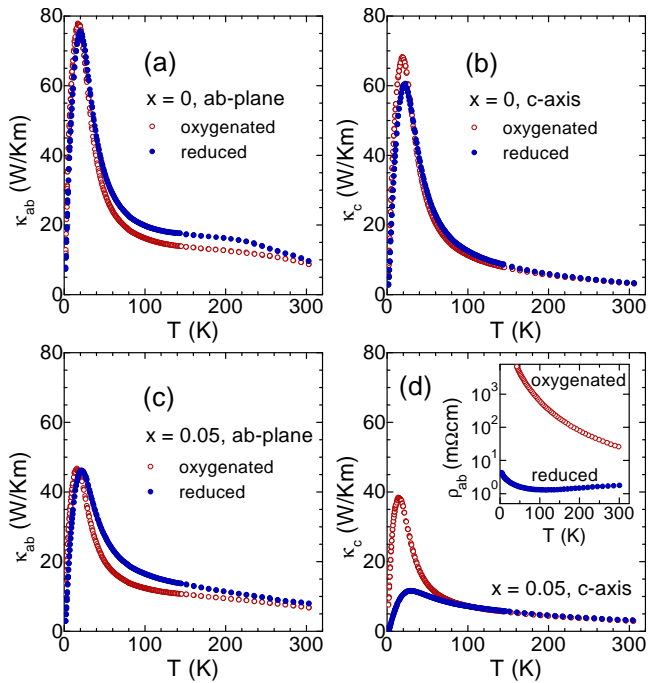} 
\caption{Comparison of the thermal conductivity of oxygenated and 
reduced Pr$_{1.3-x}$La$_{0.7}$Ce$_x$CuO$_4$ single crystals 
for (a,b) $x$ = 0 and (c,d) $x$ = 0.05. 
Inset to panel (d) shows $\rho_{ab}(T)$ data of oxygenated and 
reduced samples for $x$ = 0.05.} 
\end{figure} 

There are several possibilities for this strong doping-induced 
phonon scattering, such as the usual impurity-phonon scattering, 
electron-phonon scattering, and local structural distortions 
induced by ion substitutions. For PLCCO, thanks to the peculiar 
role of oxygen non-stoichiometry \cite{Fujita,Schultz}, a good way 
to clarify the phonon scattering mechanism is to compare the 
behavior of the reduced crystals to that of oxygenated ones, where 
the charge carrier density in the CuO$_2$ planes is depleted 
\cite{Jiang}.

Figure 3 shows such comparison for $x$ = 0 and 0.05 crystals, 
which gives us useful pieces of information: First, it is clear 
that the removal of oxygen little changes the low-$T$ phonon heat 
transport in undoped crystals [Figs. 3(a) and 3(b)], which implies 
that the apical oxygen and/or the oxygen vacancies \cite{Schultz} 
scarcely scatter phonons in both $ab$-plane and $c$-axis 
directions. Second, the peak height in $\kappa_{ab}(T)$ for $x$ = 
0.05 shows no difference between the oxygenated and reduced 
samples [Fig. 3(c)], which suggests that it is the structural 
distortion due to the Ce doping, rather than the charge carriers, 
that is mainly responsible for the scattering of the in-plane 
phonons and for the $\sim$40\% damping of the $\kappa_{ab}$ peak 
compared to $x$ = 0; this structural distortion seems to scatter 
phonons isotropically, because the $\kappa_{c}$ peak of the 
oxygenated samples is also reduced by $\sim$40\% with 0.05 Ce 
doping [see Figs. 3(b) and 3(d)]. Third, the dramatic suppression 
of the phonon peak in $\kappa_c$ with Ce doping is observed only 
in the reduced sample, which means that the strong phonon 
scattering is definitely related to the electrons in the CuO$_2$ 
planes, even though the electrons have little $c$-axis dispersion 
and are not normally expected to scatter $c$-axis phonons 
significantly. [The inset to Fig. 3(d) shows a comparison of 
$\rho_{ab}(T)$ for oxygenated and reduced $x$ = 0.05 samples, 
which demonstrates how starkly the mobile carrier density changes 
upon reduction.]

As was discussed in Ref. \cite{Sun1} for the case of LSCO, the 
lattice distortions induced by the stripes can naturally provide 
the anisotropic phonon scatterings, when the stripes are well 
ordered in the planes but are disordered along the $c$ axis 
(which is actually the case for the spin stripes in LSCO 
\cite{Matsuda}). Given the striking similarities in the 
suppression of the $\kappa_c$ peak between LSCO and PLCCO [and 
also the additional evidence in Fig. 3(d) that the strong phonon 
damping must be related to the doped electrons], it is most 
reasonable to conclude that the peculiar thermal conductivity 
behavior in PLCCO is due to the stripes that have been previously 
undetected, and the data suggest that the stripes are not 
established for $x < 0.03$. As is mentioned in the introduction, 
since the neutron scattering has found commensurate magnetic 
peaks for NCCO \cite{YamadaNCCO}, the stripes that are to be 
formed in the electron-doped cuprates should be in-phase 
antiferromagnetic domain boundaries, which do not frustrate the 
spin periodicity and are benign to the N\'eel order. We note that 
existence of the in-phase stripes has been suggested for the 
weak-ferromagnetic state of LSCO at $x$ = 0.01 under high 
magnetic field \cite{LSCO_MR}, not to mention the theoretical 
predictions for them in the literature \cite{Zachar}.

It is useful to note that some of the unusual features recently 
found in NCCO might be related to the anomalous thermal 
conductivity behavior: An optical phonon at $\sim$70 meV was 
found to soften upon slight Ce doping \cite{Kang}, and rather 
mysterious 2D superlattice structure was observed when NCCO ($x$ 
= 0.15) was reduced \cite{Kurahashi}. While it is possible that 
these features are also coming from some form of spin/charge 
texture, much remains to be sorted out about the exact roles of 
Ce doping and reduction in electron-doped cuprates. 

\begin{figure} 
\includegraphics[clip,width=8.5cm]{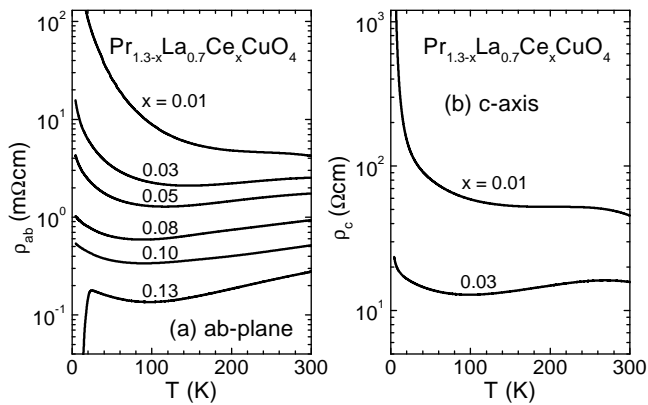} 
\caption{(a) Temperature dependences of $\rho_{ab}$ of the reduced 
single crystals of Pr$_{1.3-x}$La$_{0.7}$Ce$_x$CuO$_4$ ($x$ = 
0.01 -- 0.13) that are Ar-annealed at 850--875$^{\circ}$C. 
(b)Temperature dependences of $\rho_c$ 
of the reduced single crystals at $x$ = 0.01 and 0.03.}
\end{figure} 

In passing, we note that Hess {\it et al.} recently argued 
\cite{Hess2} that the scattering by soft phonons associated with 
the lattice instability of the low-temperature-orthorhombic (LTO) 
phase, rather than the scattering by stripes, is the source of 
the damping of the phonon peak in LSCO. However, this picture 
clearly falls short in explaining the drastic suppression of the 
$\kappa_c$ peak in the {\it lightly-doped} LSCO, where the 
LTO-transition temperature does not change much from 
La$_2$CuO$_4$ \cite{Kastner}. Furthermore, PLCCO has no LTO 
instability (this system is tetragonal irrespective of Ce doping 
and temperature) and yet shows the anomalous damping of the 
phonon peak, which gives another good reason to dismiss the soft 
phonon scenario. 

The temperature dependences of $\rho_{ab}$ are shown in Fig. 4(a) 
for $x$ = 0.01--0.13. It is intriguing to see that the $x$ = 0.01 
sample is weakly insulating ($d\rho_{ab}/dT < 0$) at room 
temperature, while the $x$ = 0.03 sample shows a metallic 
behavior; this seems to be correlated with the behavior of 
$\kappa_c$, where the anomalous damping of the low-$T$ peak 
saturates above $x$ = 0.03, and suggests that the metallic 
behavior in $\rho_{ab}$ is also due to the stripe formation. As 
was discussed for LSCO, observation of a metallic in-plane 
transport in the long-range N\'eel ordered state ($T_N$ should be 
around 200 K at $x$ = 0.03--0.08 in PLCCO \cite{Fujita}) is 
anomalous \cite{Trieste}, and it is even more peculiar that the 
hole mobility in the N\'eel state is similar to that at optimum 
doping \cite{mobility}. It was proposed that the 
self-organization of charges into stripes gives a plausible 
picture to understand this doubly-unusual metallic behavior in 
the N\'eel state \cite{mobility,Trieste}; apparently, the same 
discussion is applicable to the lightly electron-doped PLCCO, 
where the electron mobility is calculated to be around 5 
cm$^2$/Vs at 300 K, which is surprisingly similar to that in LSCO 
and is suggestive of the transport mechanism being the same. 
Therefore, the $\rho_{ab}$ data give additional support to the 
conjecture that the stripes are formed in PLCCO at $x \ge 0.03$. 
Intriguingly, the temperature dependence of the $c$-axis 
resistivity ($\rho_c$) also shows a qualitative change between 
$x$ = 0.01 and 0.03 [Fig. 4(b)].

In summary, two peculiar features in the transport properties, 
which are considered to signify the existence of stripes in the 
lightly hole-doped LSCO, are both observed in the lightly 
electron-doped PLCCO: anomalous damping of the $c$-axis phonons 
and the ``high mobility" metallic transport in the N\'eel ordered 
state. A natural conclusion of these observations is that the 
stripes are formed not only in the hole-doped cuprates but also in 
the electron-doped cuprates. Thus, there seems to be an 
approximate electron-hole symmetry for the stripe formation, 
though the detailed structure of the stripes is likely to be 
different.

We thank S. A. Kivelson, A. N. Lavrov, and J. Takeya for helpful 
discussions, and M. Fujita for giving us useful information on the 
PLCCO crystal growth.


\begin{thebibliography}{}
\bibitem{note}

The small amount of oxygen vacancies in our reduced samples 
introduce some additional electrons, which we estimate to be 
$\sim$2\% per Cu; annealing at higher temperatures 
(900--950$^{\circ}$C), which stabilizes the superconducting 
transition for $x \ge 0.10$ \cite{Fujita}, introduces further 
electron carriers and makes the $x$ value to be not a good measure 
of the electron doping.
\end{thebibliography}
\end{document}